\documentstyle[amsfonts,aps,multicol,prl,epsf]{revtex}

\newcommand{\beq}{\begin{equation}}
\newcommand{\eeq}{\end{equation}}
\newcommand{\beqa}{\begin{eqnarray}}
\newcommand{\eeqa}{\end{eqnarray}}

\begin{document}

\title{Energy landscape of a Lennard-Jones liquid: Statistics of
  stationary points}

\author{Kurt Broderix$^*$, Kamal K. Bhattacharya$^*$, Andrea Cavagna$^\dagger$, 
Annette Zippelius$^*$ and Irene Giardina$^{**}$}

\address{$^*$ Institut f\"ur Theoretische Physik, Universit\"at G\"ottingen,
Bunsenstr. 9, D-37073 G\"ottingen, Germany}
\address{$^\dagger$ 
Department of Physics and Astronomy, The University, Manchester, 
M13 9PL, United Kingdom}
\address{$^{**}$ Service de Physique Th\`eorique, CEA Saclay, 
91191 Gif-sur-Yvette, France}

\date{\today}
\maketitle

\begin{abstract}
  Molecular dynamics simulations are used to generate an ensemble of
  saddles of the potential energy of a Lennard Jones liquid.
  Classifying all extrema by their potential energy $u$ and number of
  unstable directions $k$, a well defined relation $k(u)$ is revealed.
  The degree of instability of typical stationary points vanishes at a
  threshold potential energy $u_{th}$, which lies above the energy of
  the lowest glassy minima of the system. The energies of the inherent
  states, as obtained by the Stillinger-Weber method, approach
  $u_{th}$ at a temperature close to the mode-coupling transition
  temperature $T_c$.
\end{abstract}

\begin{multicols}{2}
  
  The complex physical behaviour of supercooled liquids and glasses
  has always stimulated a description of such systems in terms of
  dynamical evolution upon a very complicated potential energy
  landscape \cite{gold}. In particular, the presence of many
  inequivalent glassy minima of the potential energy gives rise to a
  rich pattern of activated dynamics and has therefore attracted most
  of the attention of the community in the past years.  As a
  consequence, the analysis of the structure of minima has become the
  main focus of the energy landscape approach to glasses.  Based on
  the pioneering work of Goldstein \cite{gold} and of Stillinger and
  Weber \cite{sw}, a number of authors have investigated the
  statistical properties of local minima, which are reached by a
  gradient descent, starting from an equilibrium configuration of the
  supercooled liquid state at temperature $T$
  \cite{sas,gott,jon,Schroder}.  The energies of the inherent minima
  sampled after the quench reveal two characteristic temperatures: At
  low $T$ the system gets trapped in a very small number of basins and
  falls out of equilibrium. Although this transition temperature
  depends on the cooling rate, it is very close to the dynamical
  transition $T_c$ of mode-coupling theory (MCT) \cite{mct}.  A second
  higher temperature marks the onset of nonexponential relaxation,
  which has been associated with energy landscape dominated dynamics
  \cite{sas}.

A method which does not
focus purely on the properties of minima, is the instantaneous normal
mode (INM) approach \cite{inm}: the spectrum of
the eigenvalues of the Hessian matrix is computed and averaged over
all the configurations with the Boltzmann distribution at temperature
$T$.  The INM analysis focuses on two points: First, it has been
suggested that the barrier heights and hopping rates can be obtained
from the INM spectrum \cite{inm}. Second, the temperature where the 
fraction $f(T)$ of negative eigenvalues of the INM spectrum goes to zero is
interpreted as the point where the number of directions for free
diffusion in phase space vanishes, thus giving an estimate of 
$T_c$, below which activation remains as the only mechanism of
diffusion \cite{tartaglia}. 
The INM analysis has been criticized \cite{gezelter}, because equilibrium
configurations with unstable directions are in general not stationary
points of the potential energy, even if the force vanishes along the
unstable directions \cite{bl}.

In contrast to the INM method, which is an intrinsically thermal
approach by sampling configurations according to their Boltzmann
weight, we focus here on the purely geometric properties of the energy
landscape. We investigate all the stationary points of the
potential energy, be they minima, or unstable saddles \cite{ruocco}. 
We classify
them according to their number of unstable directions, (index K),
their energy and the smallest eigenvalue of the Hessian matrix.
Thereby we can address the following questions: Is there a threshold
energy for saddles, such that for energies below this threshold, it is
very unlikely to find saddles and the dynamic relaxation of the system
requires activation ?  Is there a signature of the threshold energy in
the dynamical behaviour of the system ?  What is the typical energy
difference between saddles of index $K$ and $K+1$ and can this be
taken as an estimate of the potential energy barriers ?

The system under consideration is a binary mixture of large (L) and
small (S) particles with $80\%$ large and $20\%$ small particles.
Small and large particles only differ in diameter, but have the same
mass. They interact via a Lennard--Jones potential of the form
$V_{\alpha\beta}(r_{ij})=4\,\epsilon_{\alpha\beta}
[(\sigma_{\alpha\beta}/r_{ij})^{12}
- (\sigma_{\alpha\beta}/r_{ij})^{6}]$, where $\vec{r}_i$ denotes the
position of particle i $(i=1,2...N)$ and $r_{ij}=|\vec{r}_{i}-\vec{r}_{j}|$.
All results are given in reduced
units, where $\sigma_{LL}$ was used as the length unit and
$\epsilon_{LL}$ as the energy unit. The other values of $\epsilon$ and
$\sigma$ were chosen as follows: $\epsilon_{LS} = 1.5, \sigma_{LS} =
0.8, \epsilon_{SS} = 0.5, \sigma_{SS} = 0.88$ \cite{ka}. The systems
were kept at a fixed density $\rho\approx 1.2$. Periodic boundary
conditions have been applied and the potential has been truncated
appropriately according to the minimum image rule \cite{at96}. We have
applied a truncation procedure which on the one hand ensures the
potential energy to be zero at the cut-off $r_{\rm cut}$ and on the
other hand provides a continuous first-derivative of
$V_{\alpha\beta}(r)$ at $r_{\rm cut}$.  Throughout this study we
present results for systems with $N=60$ particles using $r_{\rm cut}=1.8$.
A few samples with up to $N=120$ have been simulated to
validate our results.

In order to explore the geometric properties of the stationary points
of the potential energy, we use the following method: We start by
equilibrating a random initial configuration of $N$ particles at a
given temperature $T$ using a standard molecular dynamics simulation
technique. After equilibration the system evolves for a time
$\tau_{\rm run}$. To locate a saddle of the potential energy close 
to the equilibrated configuration we look for the absolute minima
of the modulus square of the force. In order achieve this we perform 
a quench on a pseudo-potential energy landscape $W(x)$ given by
$W(x)=\vec\nabla U(x)\cdot\vec\nabla U(x)$, whereas the original
potential energy is defined by $U(x)=\sum_{1 \leq i <j \leq N}
V_{\alpha\beta}(r_{ij})$. Notice that all {\it absolute} minima of $W(x)$
are stationary points of $U(x)$, hence every saddle of $U(x)$ has a
well defined basin of attraction. The {\it local} minima of $W(x)$
do not correspond to zeros of the real force. These points are
frequently sampled; they can easily be distinguished from the absolute
minima and are excluded from the following analysis.

Given a stationary point we consider the number of negative eigenvalues
of its Hessian matrix,
that is the index $K$ of the saddle. We compute the index
density $k=K/(3N)$ and the potential energy density $u=U/N$ of the
saddle and plot these values in the $(u,k)$ plane.  In Fig.1 we show
the results for many saddles, which were sampled with the steepest descent
procedure on the pseudo-potential $W(x)$ for two different values of
the temperature.  This plot clearly suggests that there is an
underlying curve $k(u)$, which is independent of temperature and which
encodes a purely geometric feature of the landscape.  By sampling
stationary points at different values of $T$ we are exploring
different regions of the potential energy surface and thus
different portions of the same geometric curve $k(u)$. 
\begin{figure}
\begin{center}
\leavevmode
\epsfxsize=3in
\epsffile{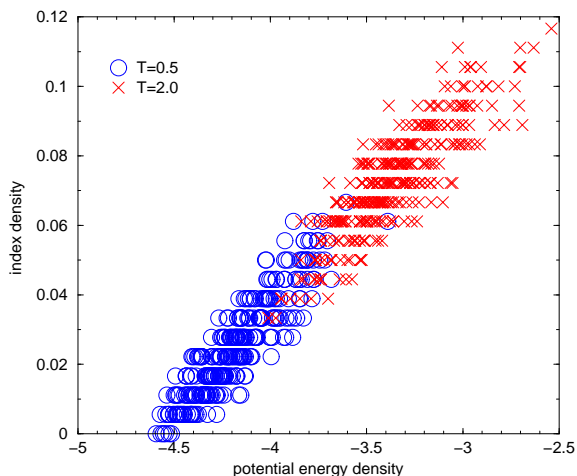}
\caption{Index density as a function of the potential energy density:
sampling of stationary points at two different temperatures, 
$T=0.5$ and $T=2.0$.} 
\end{center}
\end{figure}
\vglue -0.4 truecm
To further support this conjecture we have sampled stationary points
at various temperatures and have averaged all data in two different
ways: Firstly, we have considered all the stationary points with a
given index density and we have computed their average potential
energy (this is possible because for a finite system $k$ can only
assume discrete values). We call this procedure {\it geometric
  average}.  Secondly, we consider all the stationary points obtained
at a given temperature $T$ and we compute their average index $k(T)$
and energy $u(T)$. Eventually, we plot $k$ and $u$ parametrically in
$T$ on the $(u,k)$ plane. We call this the {\it parametric average}.
If our sampling of the saddles is a fair exploration of the underlying
geometric space, then the two averages must coincide. This is what
happens, as it is shown in Fig.2.
\begin{figure}
\begin{center}
\leavevmode
\epsfxsize=3in
\epsffile{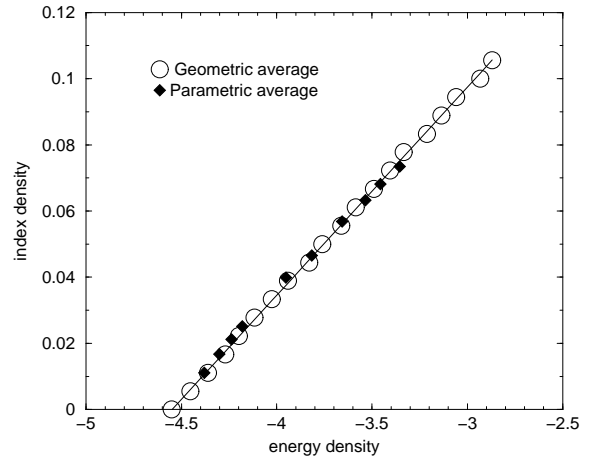}
\caption{Index density as a function of the potential energy density.
Average over all the data obtained by sampling at 
$T\in[0.3, 2.0]$. 
The full line is a linear fit of the geometric average.}
\end{center}
\end{figure}
\vglue -0.6 truecm
The averaged data as plotted in Fig.2 reveal a unique function $k(u)$.
In other words, if we cut the potential energy landscape with a plane
of constant energy density $u=u_0$, the stationary points on this
plane (or within a narrow shell around this plane) will be dominated
by saddles with  index density $k(u_0)$.
Furthermore, $k(u)$ is to a very good approximation linear up to an index
density of $10\%$ negative eigenvalues.  This implies that the curve
extrapolates to zero at a well defined energy, which we call the {\it
threshold energy}, $u_{th}$, in analogy with spin-glasses \cite{pspin}.  
In Fig.3 we present a magnification of the last four points of $k(u)$, showing
that the linear interpolation of all the data and the linear
interpolation of the last four points give the same estimate for the
threshold energy, that is $u_{th}=-4.55$.

The threshold energy marks the border between the saddles-dominated 
portion of the energy landscape and the minima-dominated one.
An interesting point is that $u_{th}$ is {\it above} the energy of the 
lowest lying minima we find, $u_0=-4.65$, as shown in Fig.3 
($u_0$ is obtained 
from an extensive search for minima of the potential energy using the 
Stillinger-Weber method). 
This implies the existence 
of a finite energy density interval where minima are overwhelmingly
more numerous than saddles. 
The same phenomenon has been observed in 
mean-field models of spin-glasses \cite{pspin} 
exhibiting one step replica symmetry breaking (1RSB).
\begin{figure}
\begin{center}
\leavevmode
\epsfxsize=3in
\epsffile{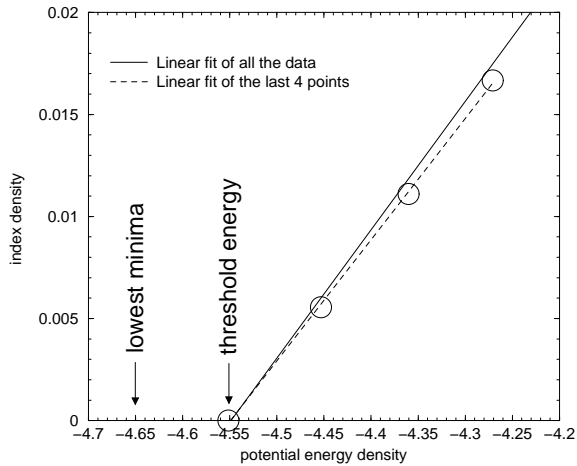}
\caption{Magnification of Fig.2. With the arrows are indicated the 
energy $u_0$ of the lowest minima found in the system and the threshold 
energy $u_{th}$.}
\end{center}
\end{figure}
\vglue -0.5 truecm
In Fig.4 we plot the lowest eigenvalue $\lambda_0$ of the Hessian 
as a function of the energy density. As expected, $\lambda_0\to 0$
for $u\to u_{th}$, implying that most extrema have a small number of 
eigenvalues close to zero.
Somewhat surprising is the approximately linear dependence of $\lambda_0(u)$
on energy, $\lambda_0\sim (u_{th} -u)$, 
exactly as in 1RSB spin-glasses \cite{pspin}. 

\begin{figure}
\begin{center}
\leavevmode
\epsfxsize=3in
\epsffile{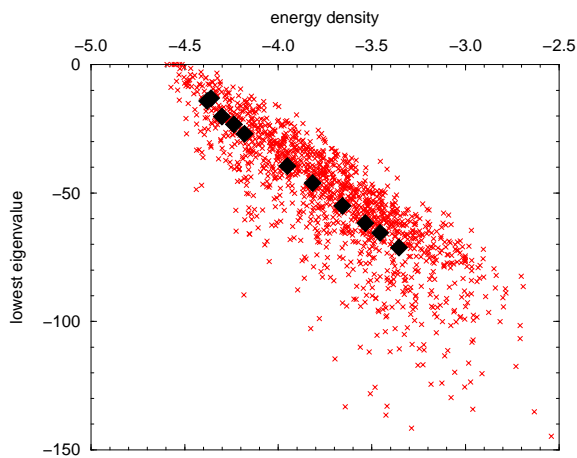}
\vskip 0.2 truecm
\caption{Lowest eigenvalue of the Hessian as a function of the energy 
density. Full diamonds are the parametric average of all the data obtained 
at the same values of $T$ as in Fig.2.}
\end{center}
\end{figure}
\vglue -0.4 truecm

It is important to relate the geometric features of the energy
landscape to the dynamical behaviour of the supercooled liquid close
to the glass transition.  The index density vanishes at the threshold
so that minima are the {\it dominant} stationary points below
$u_{th}$.  We therefore expect to find a signature of $u_{th}$ in the
dynamics or, more specifically, a link between the threshold energy
and the onset of activated dynamics upon cooling.  To that end we have
used our MD simulations at temperature T \cite{sas,gott} to compute
the energy density $u_{min}(T)$ of local minima reached by gradient
descent on the potential energy surface.  These data are compared in
Fig.5 to the difference $\delta(T)=<U/N>(T)-3T/2$ of the average
potential energy density $<U/N>(T)$, as calculated in the MD
simulation, and the vibrational energy in the harmonic approximation.
For a harmonic potential $\delta(T)$ is just the energy of the minimum
of the well. In Fig.5 we see that $\delta(T)\sim u_{min}(T)$ for
$T\leq 1.2$ \cite{sri}. This is the range of temperatures which is
dominated by the energy landscape and the timescales for the two
processes of relaxation - vibrations inside a minimum and hopping
between different minima - start to separate.  Close to the glass
transition the system falls out of equilibrium, as indicated in Fig.5
by the saturation of both quantities, $u_{min}(T)$ and $\delta(T)$.
The temperature where this happens is known to depend on the equilibration
time of the MD simulation (see e.g. \cite{sas}).
\begin{figure}
\begin{center}
\leavevmode
\epsfxsize=3in
\epsffile{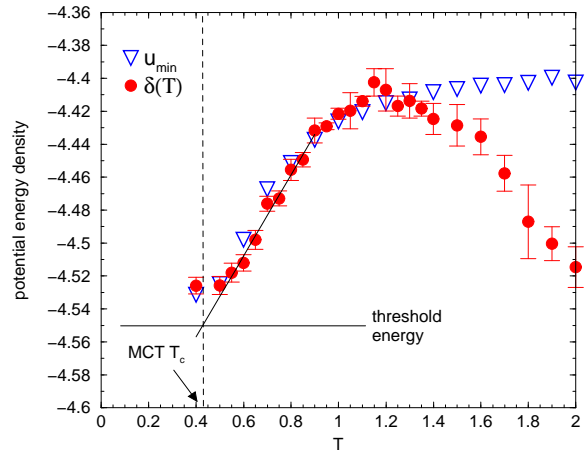}
\vskip 0.2 truecm
\caption{Triangles represent the average energy of local minima 
$u_{min}$ as a function of
the temperature of the initial MD trajectory. Circles represent the
quantity $\delta(T)=\langle U/N\rangle(T) - 3/2 T$. }
\end{center}
\end{figure}
\vglue -0.5 truecm Extrapolating that part of the curve corresponding
to temperatures where equilibration is ensured (see Fig. 5), one
observes that both $u_{min}(T)$ and $\delta(T)$ reach the threshold
energy at a temperature approximately equal to the MCT transition
$T_c\sim 0.44$\cite{ka}.  The crossover from a non-activated dynamics
above $T_c$ to an activated one below $T_c$ \cite{angell,Schroder},
may thus be interpreted in this context as a geometric transition from
saddle dominated, to minima dominated regions of the potential energy
landscape.

We can obtain an estimate of the energy barrier $\Delta U=1/(3
k^{\prime})\sim 5.0$ from the slope $k^{\prime} (u)$ of the linear
function \cite{bradipo}.  This estimate has the right order of
magnitude, see e.g. \cite{Schroder}. Note however that we do not know which
saddles are accessible from a particular minimum via a dynamic
trajectory. Furthermore we have not yet computed the entropic
contribution to the transition rate, which is necessary for an
estimate of the free energy barrier and hence for a
comparison with the experimental time scales.

Finally in Fig.6 , we compare our result for $k(u)$ to an INM
analysis. For the latter we have computed the average fraction of
negative eigenvalues for equilibrium configurations as a function of
their energy (parametric average). We see that for all 
temperatures  the INM index is higher
than the index of genuine saddles at the corresponding energy. Hence
the average curvature is overestimated by INM, as expected \cite{gezelter}.
\begin{figure}
\begin{center}
\leavevmode
\epsfxsize=3in
\epsffile{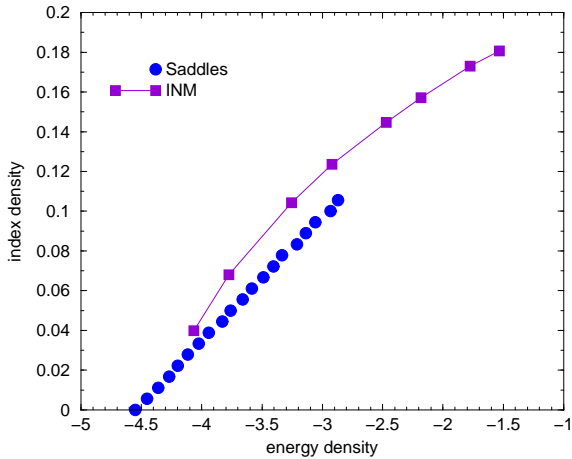}
\vskip 0.2 truecm
\caption{INM vs saddle index density as functions of the potential 
energy density. $T\in[0.3,2.0]$.}
\end{center}
\end{figure}
\vglue -0.5 truecm
In this Letter we have presented a new method to characterize 
in a quantitative way the geometric structure of the potential 
energy landscape, which goes beyond the analysis of local minima. 
We have shown that the typical saddles index $k$ 
depends linearly on the potential energy density and we
have defined a threshold energy, $u_{th}$, where the degree of 
instability vanishes. Furthermore, the threshold energy has been 
related to the MCT transition temperature $T_c$.

The method of steepest descent on the pseudo-potential $W(x)$ is
powerful enough to provide detailed information on the properties of
saddles, so that one can estimate not only energy barriers but also
the entropic contribution to the transition rates. Given that we know
all eigenvalues of the Hessian at the stationary points, we can
evaluate the partition function at the saddle, which is needed as an
input for the transition state theory. So far we have mainly looked at
one system size, i.e. $60$ particles, and have only checked a few
results for larger systems. It would be interesting to study the
effects of varying system size on barrier energies, entropies and
number of unstable directions. Another important point is the mutual
accessibility of minima and saddles. This latter problem may be
tractable by a combination of both, steepest descent on the potential
energy and steepest descent on the pseudo-potential surface.  Work
along these lines is in progress.
\vglue -0.5 truecm

\acknowledgements
A.C. thanks A. Bray, F. Colaiori, T. Keyes, M. Moore, G. Parisi, F. Ritort,
S. Sastry and F. Thalmann for useful discussions.
The work of A.C. was supported by EPSRC(UK) Grant GR/L97698 and the
work of K.B. by the DFG under grant No. Br 1894/1-1. 
\vskip 0.3 truecm
{\it In memory of Kurt}.

\vglue -0.2 truecm

\end{multicols}

\begin{thebibliography}{}

\vglue -1.0 truecm
\bibitem{gold}
M. Goldstein, J. Chem. Phys {\bf 51}, 3728 (1969).

\bibitem{sw}
F.H. Stillinger and T.A. Weber, Phys. Rev. A {\bf 25}, 978 (1982).


\bibitem{sas}
S. Sastry, P.G. Debenedetti and F. Stillinger, Nature, {\bf 393}, 554 (1998).

\bibitem{gott}
K. K. Bhattacharya, K. Broderix, R. Kree and A. Zippelius,
Europhys. Lett. {\bf 47}, 449 (1999)

\bibitem{jon} 
W. Kob, F. Sciortino and P. Tartaglia, Europhys. Lett {\bf 49}, 590 (2000)

\bibitem{Schroder}
T.B. Schr\o der, S. Sastry, J.C. Dyre and S. Glotzer, J. Chem. Phys. {\bf 112},
9834 (2000).

\bibitem{mct}
U. Bengtzelius, W. Goetze and A. Sjolander, J. Phys. Chem.  {\bf 17},
5915 (1984); 
E. Leutheusser, Phys. Rev. A {\bf 29} 2765 (1984).


\bibitem{inm} 
T. Keyes, J. Chem. Phys. A {\bf 101}, 2921 (1997).
B. Madan and T. Keyes, J. Chem. Phys. {\bf 98}, 3342 (1992).

\bibitem{tartaglia}
See, for example,
S. Sastry, Phys. Rev. Lett. {\bf 76}, 3738 (1996); F. Sciortino and 
P. Tartaglia, Phys. Rev. Lett. {\ bf 78}, 2385 (1997).


\bibitem{gezelter}
J.D. Gezelter, E. Rabani and B.J. Berne, J. Chem. Phys. {\bf 107}, 4618
(1997).

\bibitem{bl} S. Bembenek and B. Laird, Phys. Rev. Lett. {\bf 74}, 936
(1995).

\bibitem{ruocco} For a similar analysis, see also
L. Angelani, R. Di Leonardo, G. Ruocco, A. Scala and F. Sciortino,
preprint cond-mat (17 July 2000).


\bibitem{ka} W. Kob and H. C. Andersen, Phys. Rev. Lett. {\bf 73}, 1376 
(1994); Phys. Rev. E {\bf 51}, 4626 (1995)
 
\bibitem{at96} M.P. Allen and D.J. Tildesley, {\it Computer Simulations of
Liquids}, Oxford Science Publications, Oxford (1996).

\bibitem{pspin}
J. Kurchan, G. Parisi and M.A. Virasoro, J. Phys. I France {\bf 3},
1819 (1993); L.F. Cugliandolo and J. Kurchan, Phys. Rev. Lett. {\bf 71},
173 (1993); A. Cavagna, I. Giardina and G. Parisi, Phys. Rev. B {\bf 57}, 
11251 (1998).

\bibitem{sri} For an equivalent investigation, see also 
S. Sastry, J. Phys.: Condens. Matter {\bf 12}, 6515 (2000).

\bibitem{angell}
C. A. Angell, J. Phys. Chem. Sol. {\bf 49}, 863 (1988).

\bibitem{bradipo} A. Cavagna, Europhys. Lett. {\bf 53}, 490 (2001);
preprint cond-mat/9910244 (1999).

\end{thebibliography}
\end{document}